\documentclass[%
 aps,
 prb,
 amsmath,amssymb,
 showkeys,
preprint
]{revtex4-1}

\usepackage{graphicx}
\usepackage{epstopdf}
\usepackage{hyperref}

\begin{document}

\title{Electric field control of the indirect magnetic coupling through a short graphene nanoribbon}

\author{Karol~Sza{\l}owski}
\email{kszalowski@uni.lodz.pl; kszalowski@wp.pl}
\affiliation{Department of Solid State Physics, Faculty of Physics and Applied Informatics, University of {\L}\'{o}d\'{z}, ul. Pomorska 149/153, 90-236 {\L}\'{o}d\'{z}, Poland}

\date{\today}

\begin{abstract}
In the paper we consider the system composed of two magnetic planes attached to zigzag terminations of the graphene nanostructure being an ultrashort fragment of the armchair nanoribbon. We investigate theoretically an indirect coupling between these magnetic planes mediated by charge carriers as a function of external in-plane electric field. The calculations are based on a tight-binding model supplemented with Hubbard term to account for coulombic interactions. For selected sizes of the graphene nanostructure, particularly high sensitivity of the coupling to the electric field is found. This leads to the possibility of control over coupling magnitude and continuous switching of its sign between antiferromagnetic and ferromagnetic one. Such a phenomenon is demonstrated in the numerical results and its origin is analysed. The robustness of this effect against armchair edge deformation and variation of exchange energy between magnetic planes and spins of charge carriers is discussed in detailed way. 
\end{abstract}




\maketitle

\section{Introduction}

The incomparably intensive studies of graphene \cite{Novoselov1,Novoselov2} paved the way towards applications of this unique material in spintronics \cite{rmpspin,Roche2014,Pesin2012,Seneor2013}. Progress in this field involves design of novel graphene-based spintronic devices \cite{EzawaNJP,Ezawa2007,Rycerz2007,Kim2008,Spin2013,Sheng2010,Kang2012}. However, applicational potential of graphene in spin electronics is closely connected with progress of knowledge about its magnetic properties \cite{Yazyev2010}. In this context, graphene nanostructures (quantum dots, nanoflakes) attract special attention \cite{Yazyev2010,Rossier2007,Wang2008,Wang2008b,Wang2009,Zhou2011,Weymann,Weymann2,Sheng2010,Potasz2010b,Potasz2012,Ominato2013,Ominato2013b,Jaworowski2013,Kabir2014,Guo2013,Guclu2013a,Guclu2009,Guclu2011,Ezawa2007,Feldner2010,Ezawa2012,Szalowski2013c,Agapito2010}.  This is due to the fact that geometric confinement and particular form of edge can cause emergence of novel phenomena and alter substantially the magnetic characteristics predicted for infinite graphene layers (e.g. \cite{Duffy2014,Szalowski2011,Klinovaja2013,Ominato2013,Ominato2013b,Rossier2007}). 

One of the crucial factors serving the development of graphene-based spintronics is the ability to control its magnetic properties. Various approaches to manipulation of magnetism were considered in the literature, including optical methods \cite{Guclu2013a} and influence of the external magnetic field \cite{PotaszAPPA,Szalowski2013c,Guclu2013b} or hydrogenation \cite{Guo2013}. However, a fundamentally important issue in spintronic applications is controlling the magnetic properties by means of external electric field. 
Let us mention that the influence of the electric field on the magnetic properties has been noticed experimentally in some systems useful for spintronics like, for example, metallic \cite{Kawaguchi2012} and semiconducting thin films and other nanostructures \cite{Chang2013,Chiba2006,Cabral2011,Boukari2002,Owen2009,Sawicki2010,Ohno2000}. It has also been a subject of theoretical studies, among which we put particular emphasis on those devoted to modifying an indirect Ruderman-Kittel-Kasuya-Yosida interaction in the presence of electric field \cite{Zhu2010,Tamura2004}. Owing to highly unusual electronic structure, graphene and its nanostructures offer new possibilities of using the electric field to influence the energy levels available for the charge carriers and, in particular, to control the resulting magnetic properties (see the review Ref.~\onlinecite{Rozhkov2011} as well as Refs.~\onlinecite{Yun2014,Modarresi2014,Chen2013,Nair2013,Zhou2013,Santos2013a,Lee2012,Giavaras2012,Santos2013b,Parhizgar2013,Potasz2012,Munarriz2012,Potasz2010b,Raith2014,Nomura2010,Braun2011,Lu2012,Guclu2011,Agapito2010,Killi2011,Ohta2006,Jagoda2010}).

Motivated by this possibilities, we perform a theoretical study of a graphene nanostructure with two magnetic planes attached at its both ends, in search of the modifications of magnetic coupling caused by the influence of external electric field originating from the gates. Our aim is to find conditions in which the sign of the coupling can be switched between ferro- and antiferromagnetic one using the electric field. For our study we select a monolayer graphene nanostructure being a short fragment of armchair graphene nanoribbon. We mention that the magnetic properties of such structures attracted already some attention in the literature \cite{Ijas2010,Szalowski2013b,Konishi2013} and that nanoribbon-shaped structures with well-defined edges can be obtained experimentally \cite{Zhang2013,Yazyev2013,Cai2010}.

\section{Theoretical model}

\begin{figure}[t]
  \begin{center}
   \includegraphics[scale=0.9]{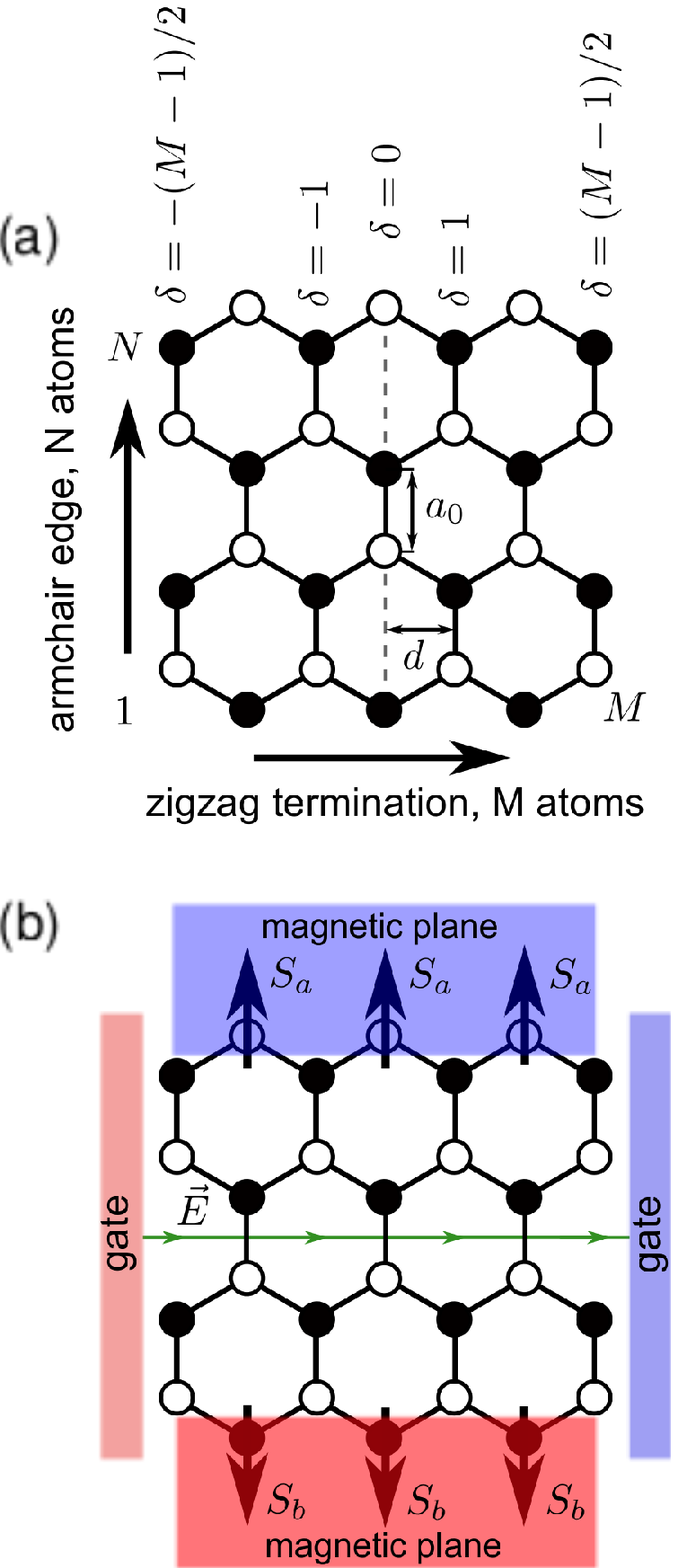}
  \end{center}
   \caption{\label{fig:fig1}(a) Schematic view of a short fragment of armchair graphene nanoribbon, characterized by zigzag terminations. The structure consists of $M$ dimer lines, each composed of $N$ atoms, where $\delta=0$ dimer line is symmetry axis. (b) View of the nanostructure placed between gates producing uniform electric field in zigzag direction and with magnetic planes attached to outermost atoms of zigzag terminations. }
\end{figure}

The subject of our interest in the present work is a graphene nanostructure in a form of a short fragment of an armchair-edged graphene nanoribbon (in fact constituting a graphene quantum dot). The schematic view of the system is presented in Fig.~\ref{fig:fig1}(a). The structure has zigzag-like terminations and is composed of $M$ atomic rows (dimer lines) extending in armchair direction. We consider only odd $M$ values, so that the structures possess symmetry axis along the dimer line in the center, the position of which we denote by $\delta=0$. The remaining dimer lines are numbered by $\delta=\pm 1,\dots,\pm (M-1)/2$. The number of atoms along each dimes line is equal to $N$, being an even number. The numbers $M$ and $N$ characterize uniquely the nanostructure, which is composed of $M\cdot N$ carbon atoms, belonging to two inequivalent, interpenetrating sublattices (as marked with filled and empty circles in Fig.~\ref{fig:fig1}). At the zigzag terminations, the magnetic planes are attached to the graphene nanostructure in such a way that the exchange interaction between magnetic moments in planes and spins of charge carriers in the nanostructure occurs only at the outermost atoms of each termination, i.e. at the ends of dimes lines with even values of $\delta$. What is more, the nanostructure is placed between the gates providing an uniform electric field along zigzag direction (perpendicular to the dimes lines). Both the magnetic planes and the gates are marked in the schematic drawing Fig.~\ref{fig:fig1}(b). 

In order to characterize the indirect magnetic interactions mediated by a graphene nanostructure, it is crucial to describe its electronic structure, in particular in the sector involving the $p^{z}$ orbitals. In our work we employ a tight-binding model supplemented with an on-site Hubbard term in mean field approximation, which is commonly used in studies of graphene nanostructures \cite{Feldner2010,Yazyev2010} (including description of indirect magnetic coupling \cite{Annika2010,Szalowski2011,SzalowskiAPPA,Szalowski2013a,Jaskolski}). The Hamiltonian of the model is the following:

\begin{align}
\label{eq:eq1}
\mathcal{H}=&-\sum_{<i,j>,\sigma}^{}{t_{ij}\,\left(c^{\dagger}_{i,\sigma}c_{j,\sigma}+c^{\dagger}_{j,\sigma}c_{i,\sigma}\right)}\nonumber\\&+U\sum_{i}^{}{\left(\left\langle n_{i,\uparrow}\right\rangle n_{i,\downarrow}+\left\langle n_{i,\downarrow}\right\rangle n_{i,\uparrow}\right)}\nonumber\\&-U\sum_{i}^{}{\left\langle n_{i,\uparrow}\right\rangle\left\langle n_{i,\downarrow}\right\rangle}+eEd\sum_{i,\sigma}^{}{n_{i,\sigma}\delta_{i}}\nonumber\\&+\frac{J}{2}S_{a}\sum_{a}^{}{\left(n_{a,\uparrow}-n_{a,\downarrow}\right)}+\frac{J}{2}S_{b}\sum_{b}^{}{\left(n_{b,\uparrow}-n_{b,\downarrow}\right)}.
\end{align}

The operator $c^{\dagger}_{i,\sigma}$ ($c_{i,\sigma}$) creates (annihilates) an electron with spin $\sigma=\uparrow,\downarrow$ at site $i$, while $n_{i,\sigma}=c^{\dagger}_{i,\sigma}c_{i,\sigma}$ represents the number of electrons. In tight-binding term, $t_{ij}$ is the hopping integral between nearest-neighbour carbon atoms belonging to a pair $\left\langle i,j\right\rangle$. For atomic pairs at the armchair edges of the nanostructure (i.e. for $\delta=\pm(M-1)/2$), its value is taken as $t_{ij}=t\left(1+\Delta\right)$. This selection depicts the edge deformation of the armchair graphene nanoribbons consisting in shortening of the interatomic distances along the armchair edges \cite{Son2006}; we use the value of $\Delta=0.12$ (what corresponds to bond deformation of approximately 3.5 \% \cite{Son2006}). For the rest of the nearest-neighbour pairs, we assume $t_{ij}=t$. The parameter $t$ is taken in further calculations as a convenient energy scale, to which all the other quantities can be normalized. The value of $t$ amounts to approximately 2.8 eV \cite{rmp} in graphene. The on-site Hubbard repulsion energy $U$ is an effective parameter, whose value takes into consideration the long-range nature of coulombic interactions and is estimated as $U/t=1$ \cite{Schuler2013}, which value we accept in our calculations. The external uniform electric field of the gates is denoted by $E$, while $d=a_{0}\sqrt{3}/2$ is the distance between the subsequent parallel atomic rows in armchair direction (i.e. distance between the dimer lines) and $\delta_{i}=0,\pm 1,\dots,\pm (M-1)/2$ is the number of dimer line to which $i$-th atom belongs. The distance between nearest-neighbour carbon atoms we denote by $a_0$ (which is about 1.4 \AA \cite{rmp}). It is instructive to mention that the normalized field of $Ed/t=1$ corresponds to the electric field of approximately 2.3 V/\AA. The interaction between the spins of charge carriers and magnetic moments of both magnetic planes attached to the zigzag terminations of the nanostructure is parametrized by exchange energy $J$. As mentioned, the coupling involves only the spins of the charge carriers present at zigzag terminations for even values of $\delta$ (and the number of such sites at each termination is $(M-1)/2$). Therefore, $a$ and $b$ denote only the summation over the outermost lattice sites of zigzag terminations. The magnetic moments associated with the planes interacting with charge carrier spins possess magnitude of $S$ and are denoted by $S_a$ and $S_b$, respectively. Let us state that the problem of selection of the value of exchange energy between magnetic impurity spin and charge carrier spin in graphene is present also in other theoretical works, for example those related to spin relaxation in graphene \cite{Gmitra2014}, where the value of $J=0.4$ eV was selected as representative. On the other hand, a value of $J=1.0$ eV was used in interpretation of experimental results on spin current scattering \cite{McCreary2009}. 

The Hamiltonian (Eq.~\ref{eq:eq1}) can be presented in a form of $\displaystyle\mathcal{H}=\mathcal{H}_{\uparrow}+\mathcal{H}_{\downarrow}-U\sum_{i}^{}{\left\langle n_{i,\uparrow}\right\rangle\left\langle n_{i,\downarrow}\right\rangle}$, with two interdependent Hamiltonians $\mathcal{H}_{\uparrow}$, $\mathcal{H}_{\downarrow}$ for spin up and spin-down charge carriers (see e.g. the detailed description in Ref.~\cite{Maiti2010}). The presence of $N_e=M\cdot N$ electrons on $p^{z}$ orbitals is assumed, what corresponds to charge neutrality of the structure (one electron per carbon atom). Both Hamiltonians can be diagonalized numerically in single-particle approximation, what allows for self-consistent determination of the energy eigenvalues  $\epsilon_{j,\sigma}$ for both orientations of spin. As a consequence, the total energy of the charge carriers in the ground state can be calculated as $\displaystyle E=\sum_{j=1}^{N_{e,\uparrow}}{\epsilon_{j,\uparrow}}+\sum_{j=1}^{N_{e,\downarrow}}{\epsilon_{j,\downarrow}}-U\sum_{i}^{}{\left\langle n_{i,\uparrow}\right\rangle\left\langle n_{i,\downarrow}\right\rangle}$, where average values of charge concentrations $\left\langle n\right\rangle$ correspond to self-consistent solution and the single-particle eigenenergies are assumed to be sorted in ascending order. In order to reach a ground state of the system, the self-consistent calculations are performed separately for all the possible numbers of electrons with spin up ($N_{e,\uparrow}$) and spin down ($N_{e,\downarrow}$) satisfying $N_{e,\uparrow}+N_{e,\downarrow}=N_e$ and the state with the lowest total energy is accepted. Moreover, the calculations are repeated for random initial conditions (charge distribution) for the purpose of finding a true ground state. 

If diagonalization of is performed both for antiparallel and parallel orientation of spins of magnetic planes $S_a$ and $S_b$, then an indirect coupling energy between the planes can be determined. It is expressed by the formula $2J^{indirect}S^2=E^{AF}-E^{F}$, where $E^{F}$ ($E^{AF}$) is the total energy of charge carriers for parallel (antiparallel) orientation of spins $S_{a}$ and $S_{b}$\cite{Annika2010,Szalowski2011,Szalowski2013a}. Let us put emphasis on the fact that such a calculation bases on a non-perturbative approach. In our work the numerical diagonalization of the Hamiltonian (Eq.~\ref{eq:eq1}) was performed with the help of the LAPACK package \cite{lapack}. 

The presented model allows to study the influence of the external in-plane electric field on the indirect, graphene nanostructure-mediated coupling between magnetic planes for various sizes $M$ and $N$. This enables identification of the most interesting cases for which the variation of coupling under the influence of the electric field is most significant. In our analysis we focus on rather small structures, composed of a few tens of carbon atoms. What is more, due to the the non-perturbative nature of the calculations, a non-trivial analysis of the influence of exchange energy $J$ on the resulting $J^{indirect}$ is possible, thus various contributions to coupling can be identified.

\section{Results}

\begin{figure}[t]
  \begin{center}
   \includegraphics[scale=0.3]{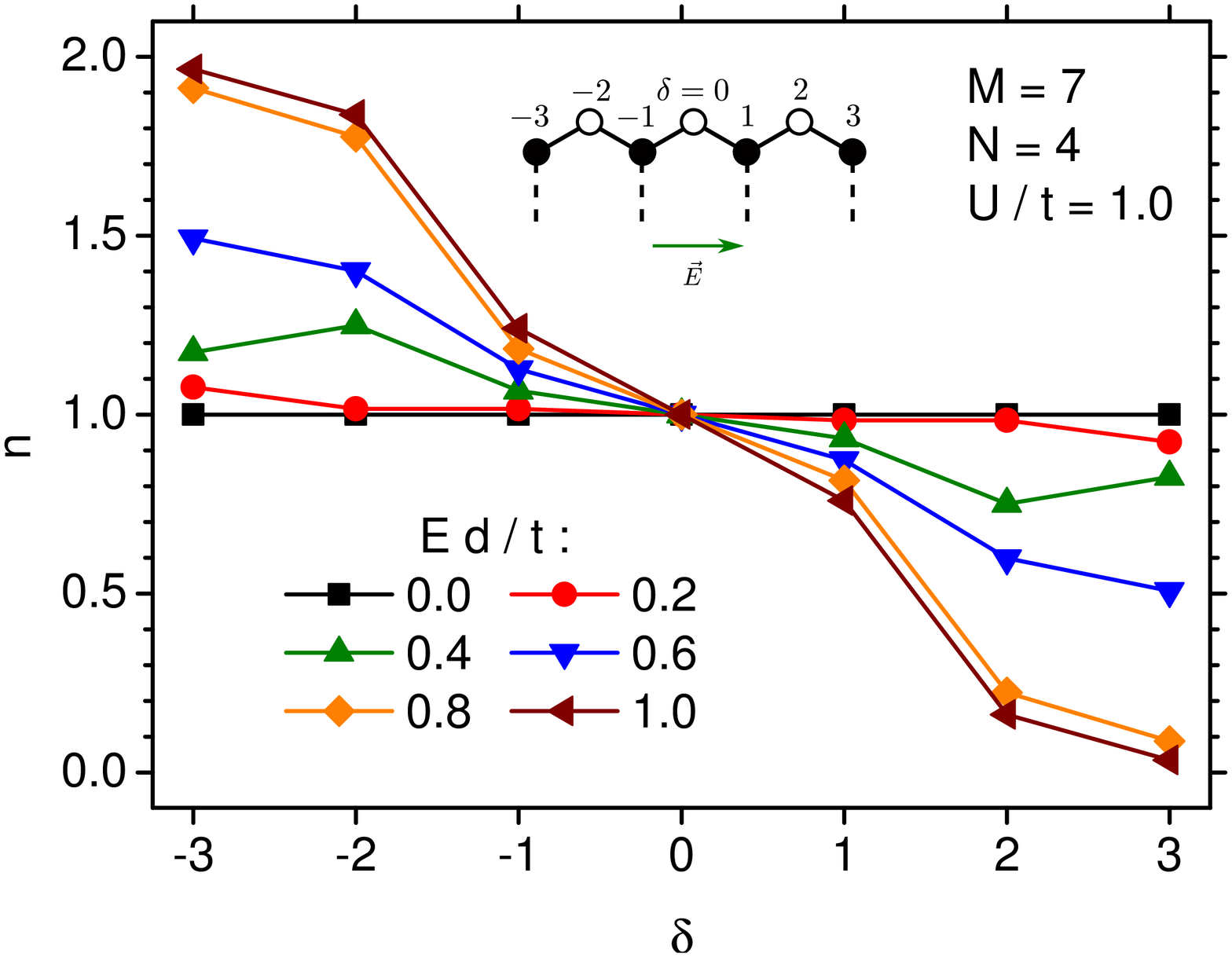}
  \end{center}
   \caption{\label{fig:fig2} Charge density at the lattice sites located at zigzag terminations of nanostructure with $M=7$ and $N=4$ for various strengths of external electric field.}
\end{figure}

At the beginning it is instructive to study some properties of the graphene nanostructures in question without including the interaction with magnetic planes. This involves in particular the influence of the electric field on the charge and spin distribution. Since the magnetic planes are attached to the outermost parts of the zigzag terminations of the structures (to the lattice sites with even values of $\delta$), it is most interesting to analyse the concentration of charge carriers particularly at these lattice sites. The distribution of total charge density $n$ (for both spin-up and spin-down electrons) along the zigzag termination of the nanostructure is illustrated in Fig.~\ref{fig:fig2} for various values of external electric field in a nanostructure characterized by $M=7$ and $N=4$. For $E=0$ the distribution of charge is homogeneous. When in-plane electric field parallel to the zigzag terminations of the nanostructure is switched on, the charge redistribution starts. The charge density at the symmetry axis ($\delta=0$) remains insensitive to the field, while the most pronounced changes occur at the edges of the nanostructure ($\delta=\pm 3$), where they are first observable for weak field. Close to the field of $Ed/t=1.0$, we deal with almost linear dependence of charge density on distance, as the edge near the negative potential gate is almost completely free from electrons, while at the opposite side, near the positive potential gate the electronic density is almost doubled. It is visible that the changes of electronic density are non-linear in external field $E$, unless the lattice sites lie close to the symmetry axis. 

\begin{figure}[t]
  \begin{center}
   \includegraphics[scale=0.3]{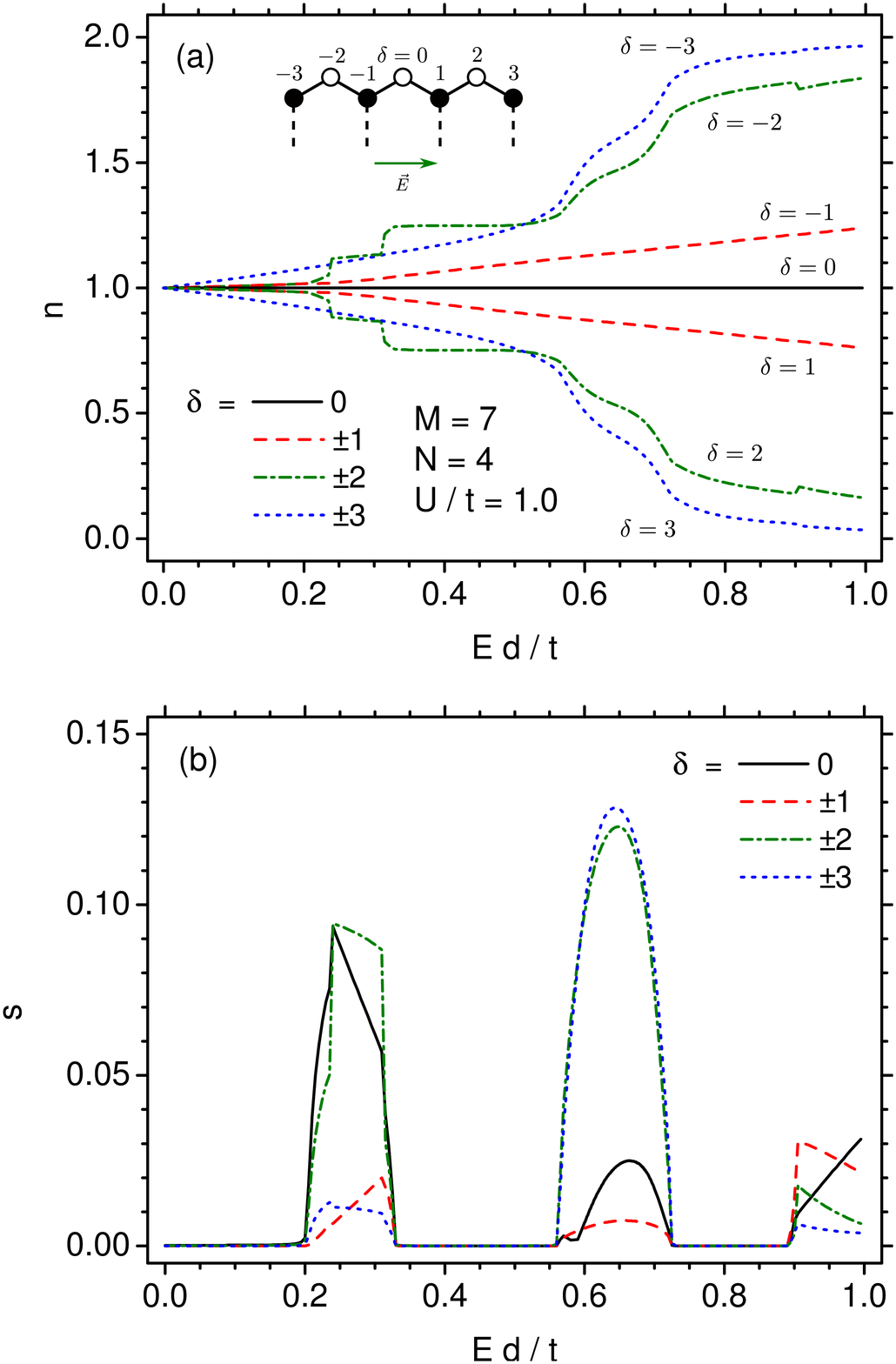}
  \end{center}
   \caption{\label{fig:fig3}(a) Dependence of (a) charge density and (b) spin density at lattice sites on the zigzag termination of the nanostructure [see inset in (a)] on the normalized electric field. Both plots are prepared for the structure with $M=7$ and $N=4$.}
\end{figure}

This feature can be followed in the dependence of the total electronic density $n$ at selected carbon sites on the in-plane electric field, which is shown in Fig.~\ref{fig:fig3}(a) (for the nanostructure with $M=7$ and $N=4$). The data concern the sites at the nanostructure terminations, where the charge carrier spins interact with magnetic planes. Such sites are marked in the inset to Fig.~\ref{fig:fig3}(a). It is visible that at the site located at the symmetry axis of the nanoribbon, for $\delta=0$, the electronic density is completely untouched by the external electric field. For the sites closer to the gate with negative potential the electronic density deceases, while the opposite changes occur at the sites located closer to the gate possessing positive potential. For the sites located at $\pm\delta$, the changes are symmetric with respect to the value of $n=1$. When the sites located closest to the edge ($\delta=\pm 3$) are considered, the dependence remains linear up to $Ed/t\simeq 0.5$, and then abrupt jumps of electronic density occur. For the field approaching $Ed/t=1.0$, electronic densities at those edge sites reach the saturation values (equal to 0 for the site close to the negative potential gate and equal to 2 for the site closest to the positive potential gate). For the sites located at $\delta=\pm 2$, the initial slope of the dependence is greatly reduced, but the abrupt changes in $n$ occur at weaker field, first of them approximately at $Ed/t\simeq 0.2$. High-field jumps take place at similar values as for $\delta=\pm 3$, but the saturation values are already not reached at $Ed/t=1.0$. For the sites closest to the symmetry axis ($\delta=\pm 1$), the initial dependence of $n$ on $E$ is linear with small slope, and this slope increases and reaches a constant value at about $Ed/t\simeq 0.3$.

\begin{figure}[t]
  \begin{center}
\includegraphics[scale=0.4]{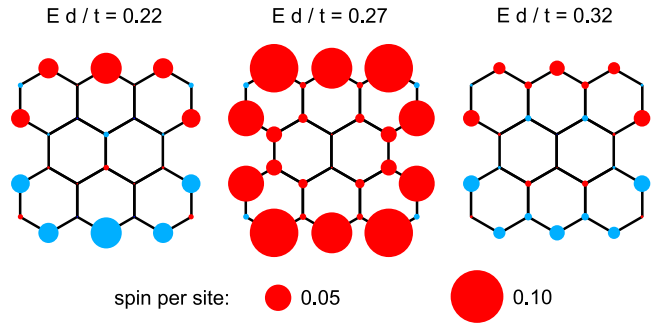}
  \end{center}
   \caption{\label{fig:fig3b}Spin density distribution for the nanostructure with with $M=7$ and $N=4$, in absence of interaction with external magnetic planes, for various electric fields for which nonzero spin at zigzag terminations emerges. Two different colours correspond to opposite orientations of electron spins. }
\end{figure}

\begin{figure}[t]
  \begin{center}
\includegraphics[scale=0.3]{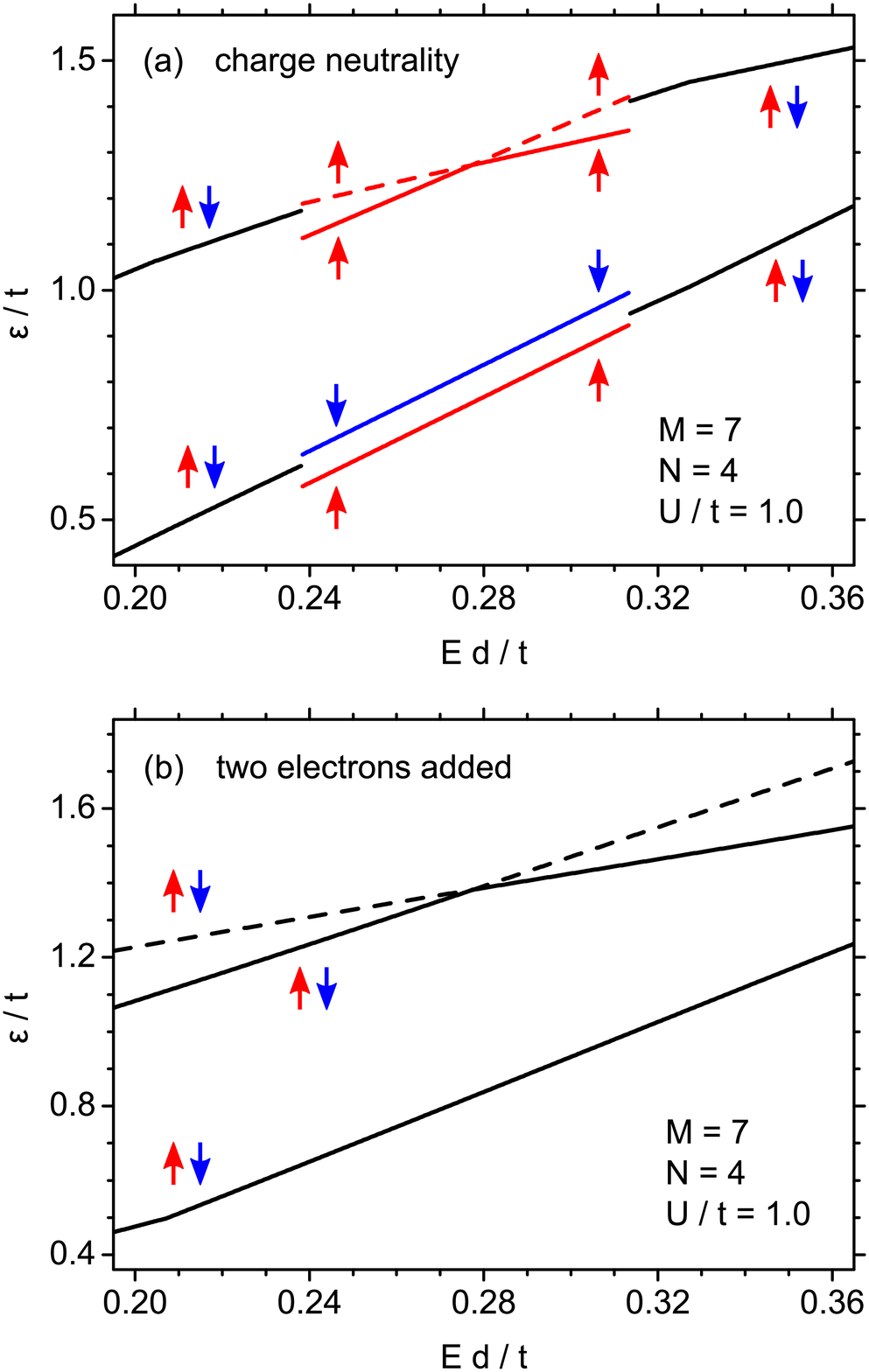}
  \end{center}
   \caption{\label{fig:fignew}Energies of a few occupied electronic states close to the highest occupied one as a function of normalized electric field, for the nanostructure with with $M=7$ and $N=4$, in absence of interaction with external magnetic planes. The case (a) is for charge neutrality, while case (b) is for two electrons added to the system. }
\end{figure}

This picture can be supplemented with the dependence of spin density $s=\left(n_{\uparrow}-n_{\downarrow}\right)/2$ on the electric field, presented in Fig.~\ref{fig:fig3}(b) for the same sites as in Fig.~\ref{fig:fig3}(a). The spin densities at sites $\pm\delta$ are identical, thus full symmetry with respect to the nanostructure axis $\delta=0$ is present. Up to $Ed/t\simeq 0.2$ there is no spin polarization at the considered sites. However, at this field value, a range of nonzero spin polarization builds up and preserves up to $Ed/t\simeq 0.33$. This happens simultaneously for all the sites, however, the values of spin polarization are much higher for $\delta=0$ and $\delta=\pm 2$ than for the sites numbered by odd $\delta$ values. This observation is particularly important in context of the fact that the magnetic planes are attached to the sites numbered by even values of $\delta$. Another range of nonzero $s$ is present between $Ed/t\simeq 0.55$ and $Ed/t\simeq 0.73$ (that time the polarization is significantly stronger close to the edges than close to the symmetry axis of the nanostructure). The next range of nonvanishing polarization commences at the field of $Ed/t\simeq 0.9$. Let us comment that these results do not imply necessarily appearance of a nonzero total spin of the nanostructure. Such a polarization does not take place in absence of electric field for the systems in question, what is due to the fact that the nanoflakes contain equal number of carbon atoms from both sublattices and no net spin polarization is predicted for charge neutrality conditions in such nanostructures \cite{Yazyev2010}. However, in some range of the electric fields (for example $0.25\lesssim Ed/t\lesssim 0.31$ and $Ed/t\gtrsim 0.90$), the total spin of the nanostructure equal to 1 is predicted. In order to illustrate that, we present the spin density distribution for a whole nanostructure in Fig.~\ref{fig:fig3b}, for three values of electric field. All three values fall within the range of appearance of nonzero spin polarization of the zigzag termination. It is visible that for the largest and the smallest value (i.e. for $Ed/t=0.22$ and $0.32$), the total spin of the charge carriers in the nanostructure is zero, with opposite spin polarization of both edges. On the contrary, for $Ed/t=0.27$, a nonzero total spin emerges. All the predicted spin polarizations manifest themselves mostly at the outermost sites of the nanostructure, mainly its zigzag terminations.

In order to gain more insight into physical mechanism leading to the polarization, we studied the energies of the discrete electronic states as a function of the electric field. In particular, we focused our attention on the lowest field range in which magnetic polarization appears in the nanostructure with $M=7$ and $N=4$. The results are shown in Fig.~\ref{fig:fignew}, where the energies of a few occupied states close to the highest occupied one are plotted against normalized electric field. Fig.~\ref{fig:fignew}(a) corresponds to the case of charge-undoped nanoribbon, while Fig.~\ref{fig:fignew}(b) is prepared for a nanostructure doped with two additional electrons. For the system with two additional electrons, no spin polarization emerges due to external electric field and all the states are spin-degenerate (this case is shown as a reference one). However, in a charge-neutral system, the situation is different. For the state with the lowest energy shown in Fig.~\ref{fig:fignew}(a), the Zeeman splitting occurs in some $E$ range, so that spin-degeneracy is lifted. However, the picture is slightly more complex in the case of the state with higher energy (the highest occupied state). Namely, it lies in the vicinity of another state (the one plotted with dashed line in Fig.~\ref{fig:fignew}(b). When both states experience Zeeman splitting, it becomes more energetically favourable for one of the electrons to skip to the next available state (plotted with dashed line both in Fig.~\ref{fig:fignew}(a) and Fig.~\ref{fig:fignew}(b)). Such a situation takes place because the energy of that state for spin up becomes lower that the energy of the previous state for spin down. Therefore, net polarization with spin equal to 1 occurs. It can be concluded from the analysis of the Fig.~\ref{fig:fignew} that the appearance of the spin polarization is associated with a single electronic state and requires a pair of states with small separation in energy.

Let us now turn the attention to the question of size of the graphene nanostructure mediating the interaction between the magnetic planes. Extending the mediating graphene nanostructure would lead eventually to achieving a limit of an infinite graphene monolayer, which is actually a zero-gap semiconductor and conducts current. Therefore, under such conditions, the model Hamiltonian which we employ to describe the influence of the electric field on the system would no longer be physically justified. On the contrary, the applicability of the model is limited to systems with well-separated discrete energy states (especially with pronounced highest occupied molecular orbital-lowest unoccupied molecular orbital gap). In particular, the spatial extension of the mediating nanostructure in the direction parallel to the electric field should be strongly limited due to screening effects. Let us mention here the results of calculations in Ref.~\onlinecite{Yamanaka2014}, concerning the screening of the parallel electric field in graphene nanostructures of various chirality. It has been found that the screening appears the least pronounced and the potential profile across the structure is most similar to linear one in the case of an amchair nanoribbon (see Fig.~3(a) in Ref.~\onlinecite{Yamanaka2014}). However, increasing the width of the structure would certainly limit the validity of the model which we employ to describe the effect.

Another factor enforcing the usage of small graphene nanostructures is the fact that the switching of the indirect coupling is vitally dependent on the behaviour of a single energy state under the influence of the electric field, as explained before. Such a state develops spin polarization and spin density associated with individual lattice sites decreases with the increase of the number of carbon atoms building the structure. Therefore, the spin density at each outermost site of zigzag terminations, being in contact with magnetic plane, would be significantly limited in excessively large structures. This limitation applies to the possibility of increasing the length of the nanostructure (i.e. the distance between magnetic planes). On the other hand, limitations for increasing the width of the nanostructure have been discussed in the above paragraph. 

These considerations are aimed at justifying the interest in the smallest graphene nanostructures in the present paper.

The creation of nonuniform spin polarization at the zigzag termination of the structure under the influence of the electric field is particularly useful from the point of view of achieving electric-field control over the indirect coupling by attaching the magnetic planes to the edges. Since the outermost lattice sites at both zigzag terminations belong to different sublattices (see Fig.~\ref{fig:fig1}), without electric field the indirect coupling is expected to indicate antiferromagnetic character (e.g. \cite{Annika2010,Crystals}). This kind of coupling results from typical indirect RKKY mechanism, which consists in creation of spin polarization by one of the magnetic planes and the interaction of the other plane with such polarization. Therefore, this process is describable (for low $J$ exchange energies) by means of second order perturbation calculus. However, the presence of the electric-field induced spin polarization at the lattice sites where charge carrier spins interact with magnetic moments in the attached planes opens the room for appearance of another coupling mechanism. Namely, the first-order perturbative coupling mediated by spin-polarized states can become active. Such a coupling mechanism has already been predicted mainly in graphene nanoflakes \cite{Szalowski2011,Szalowski2013a,SzalowskiAPPA}, but also in armchair graphene nanoribbons of finite length \cite{Szalowski2013b}. Therefore, the described effect of electric field on the electronic properties of the studied nanostructures potentially allow for continuous changing the indirect interaction strength and sign, which property we will show in further part of this section.

\begin{figure}[t]
  \begin{center}
   \includegraphics[scale=0.3]{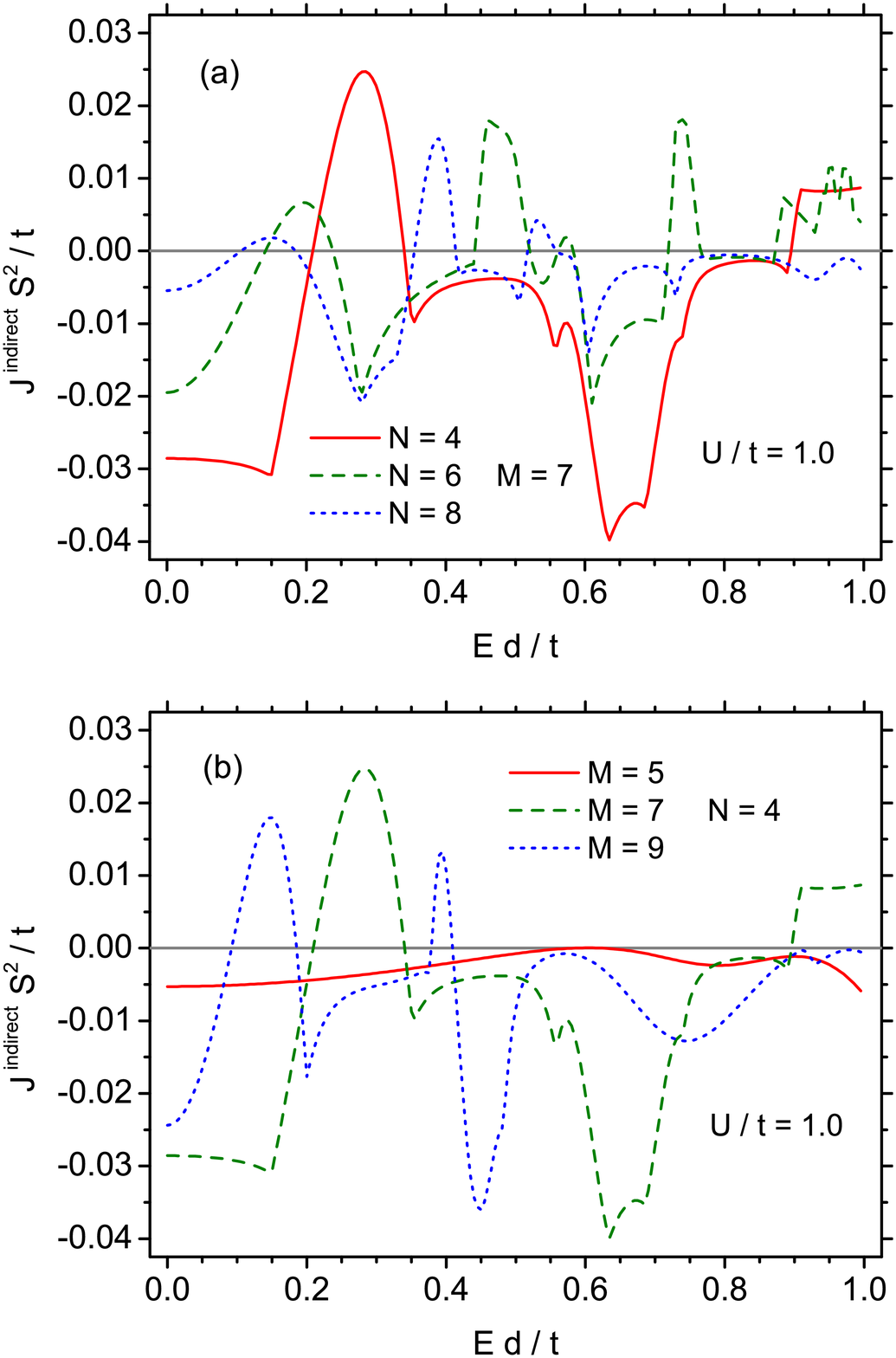}
  \end{center}
   \caption{\label{fig:fig4}Normalized indirect coupling as a function of normalized electric field in nanostructures of varying size: (a) for width $M=7$ and three lengths: $N=4,6,8$; (b) for length $N=4$ and three widths: $M=5,7,9$.}
\end{figure}

Let us now present the results of calculations for the complete system, including the magnetic planes. First we can follow the dependence of indirect coupling energy between the magnetic planes attached to the zigzag terminations of the sample graphene nanostructure as a function of the external electric field. The plot Fig.~\ref{fig:fig4}(a) presents the predictions of calculations for three representative nanostructures, each of width $M=7$, but with lengths equal to $N=4$, $6$ and $8$. The exchange energy is taken as $J/t=0.1$ and armchair edge deformation is included. First it is visible that for low (or zero) electric field the coupling is antiferromagnetic (AF) and its magnitude decreases with the increase of the structure size. When the field increases, $J^{coupling}$ indicates first some plateau, but then becomes weaker and passes through zero, changing its sign to ferromagnetic (F). This is the effect we expect on the basis of the analysis of spin polarization. For the nanostructure with $M=7$ and $N=4$, the field at which the sign change occurs coincides with the onset of spin polarization visible in Fig.~\ref{fig:fig3}(a). The coupling energy reaches a maximum and the drops again to antiferromagnetic values when $E$ increases further. A qualitatively similar behaviour of $J^{coupling}$ as a function of electric field is seen for the longer nanostructures, with $N=6$ and $N=8$. However, it that cases the ferromagnetic maximum is much reduced in magnitude and is shifted to weaker fields $E$. What is more, the oscillatory character of changes of $J^{indirect}$ for higher fields becomes more pronounced for larger structures. Finally, let us observe that for larger lengths the low-field ferromagnetic maximum tends to vanish.

An analogous analysis is presented in Fig.~\ref{fig:fig4}(b), but that time for structures sharing the same length $N=4$ and different in width: $M=5$, $7$ or $9$. For the narrowest of the nanostructures, the magnitudes of indirect coupling are generally reduced and for increasing $E$ the interaction energy slowly tends to zero, however, without visible ferromagnetic maximum. When $M$ increases to $7$, a behaviour described in the above paragraph emerges. If $M=9$, still a ferromagnetic maximum occurs (like that for $M=7$), but the strongest coupling is reduced and the peak is present at weaker fields.

\begin{figure}[t]
  \begin{center}
   \includegraphics[scale=0.3]{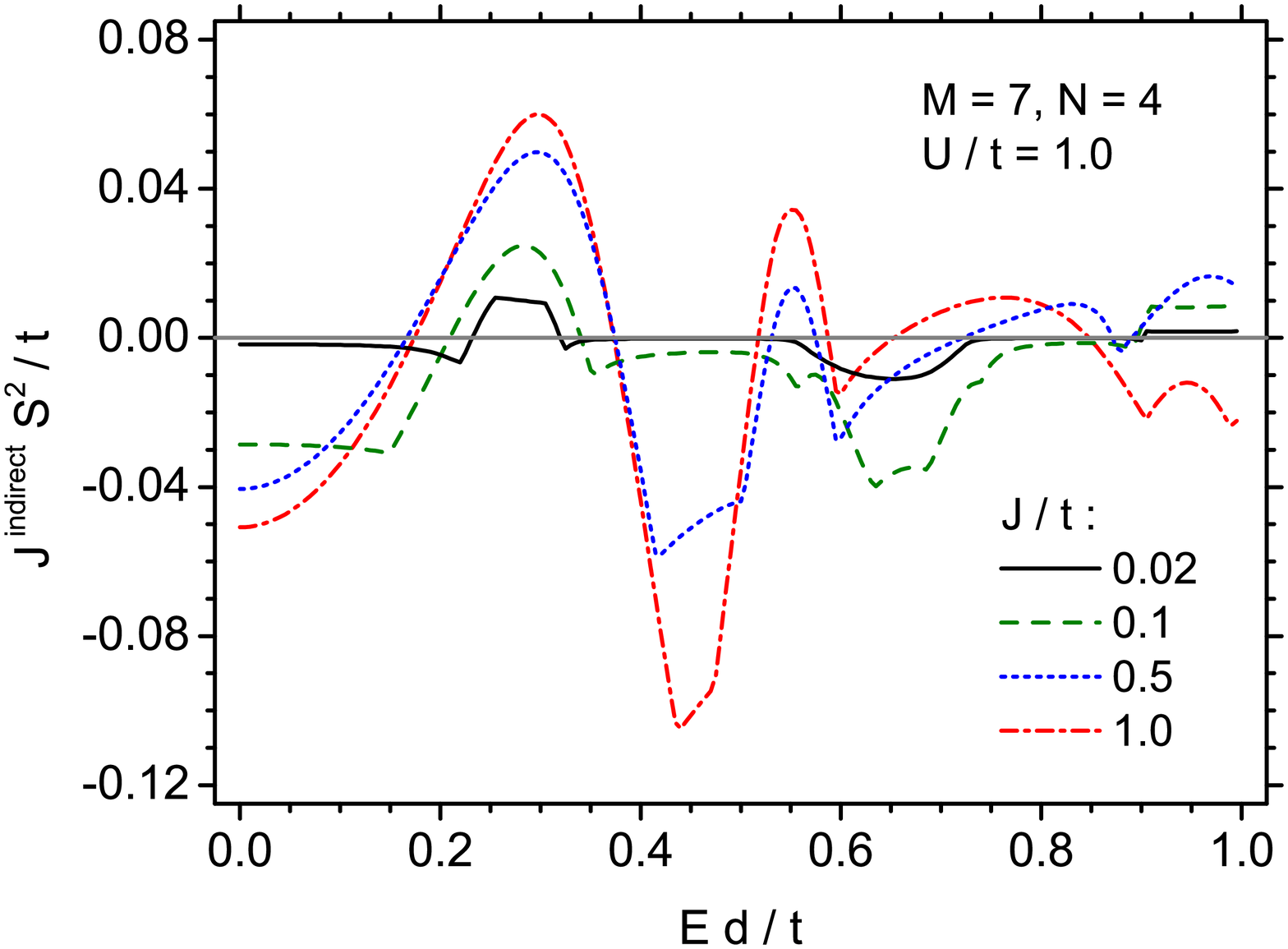}
  \end{center}
   \caption{\label{fig:fig5} Normalized indirect coupling as a function of normalized electric field in nanostructure characterized by $M=7$ and $N=4$, for four different values of exchange energy $J$.}
\end{figure}

The results described above allow to state that for some sizes of the studied nanostructures, an electric field-induced, continuous switching between antiferro- and ferromagnetic indirect coupling can be achieved. It is further interesting to verify the stability of the effect with respect to the value of exchange energy parameter $J$. Dependencies of $J^{indirect}$ on electric field $E$ are plotted in Fig.~\ref{fig:fig5} for four selected values of exchange energy $J$, ranging from $J/t=0.02$ up to $J/t=1.0$. The selected nanostructure is the one which provides the most pronounced field-induced ferromagnetic peak in indirect interaction, i.e. $M=7$ and $N=4$. It is visible that the ferromagnetic peak occurs for all the values of $J/t$ and its location is similar in all the cases. On the other hand, for the antiferromagnetic ranges surrounding this peak, $J^{indirect}$ is very sensitive to exchange coupling energy $J$. Especially, a very deep minimum can emerge at around $E/t\simeq 0.45$ when $J/t$ is strong enough. The values of the electric field at which $J^{indirect}$ crosses zero (the boundaries of the ferromagnetic range) can be studied in more detailed way as a function of exchange energy $J/t$ for various sizes of the nanostructure.

\begin{figure}[t]
  \begin{center}
   \includegraphics[scale=0.3]{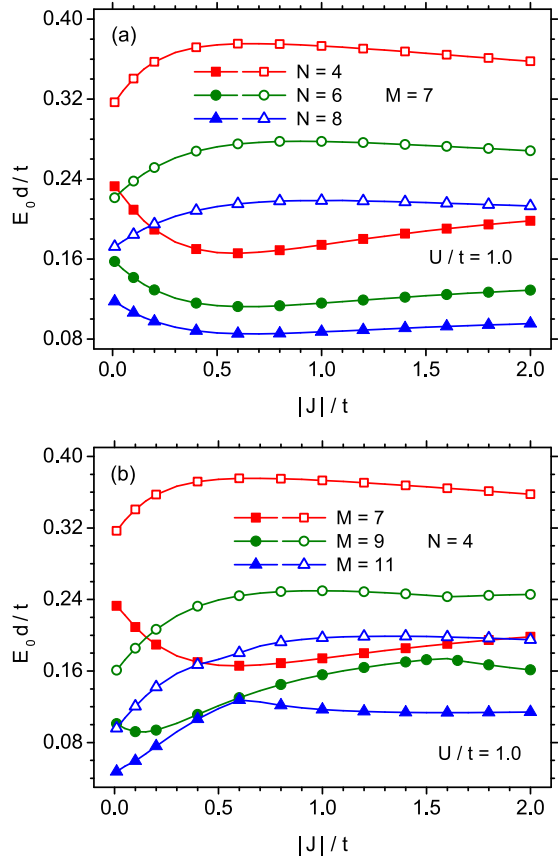}
  \end{center}
   \caption{\label{fig:fig9}Critical normalized electric field of transition between ferro- and antiferromagnetic coupling between magnetic planes as a function of exchange energy $J$ in nanostructures of varying size: (a) for width $M=7$ and three lengths: $N=4,6,8$; (b) for length $N=4$ and three widths: $N=5,7,9$. Filled symbols denote the lower field, at which AF coupling turns into AF one, while filled symbols correspond to a higher field at which opposite transition takes place. }
\end{figure}

In view of the potentially wide range of parameters $J$ describing the exchange coupling between magnetic planes and spins of the charge carriers, it is highly interesting to determine the behaviour of the critical electric field $E_0$ for which the indirect coupling crosses the zero value as a function of exchange energy $J/t$. Such results are presented in Fig.~\ref{fig:fig9} for various sizes of a nanoribbon piece mediating the coupling. Filled symbols correspond to the field of transition between low-field AF interaction and higher-field F interaction. On the other hand, empty symbols denote the electric field at which F-interaction switches back to AF, what takes place for higher fields. In Fig.~\ref{fig:fig9}(a) structures of width $M=7$ and three lengths: $N=4,6,8$ are considered. It is visible that the dependence of the normalized electric fields $E_0 d/t$ on $J/t$ is not very pronounced (unless $J/t\lesssim 0.3$), the main feature being a shallow extremum in the vicinity of $J/t\simeq 0.6$. Therefore, around this value of $J$ the ferromagnetic interaction covers the widest range of electric fields. The tendencies of variation of both characteristic fields (for AF-F and F-AF transition) as a function of $J$ are opposite. What is crucial, the effect of changing the coupling $J^{indirect}$ from AF to F is present in the whole studied range of exchange potentials $J$. The situation is slightly different in Fig.~\ref{fig:fig9}(b), which presents analogous data for structures with constant length $N=4$ and three different widths $M=7,9,11$. When the width increases, the minimum in dependence of $E_0$ for AF-F transition on $J$ shifts significantly to lower values of $J$. Moreover, for $N>7$ an additional maximum appears (also moving to lower $J/t$ when $N$ increases). The overall dynamics of the critical fields $E_0$ is more significant for wider pieces of nanoribbons. The field of F-AF transition (the higher one) shows in this case a very similar behaviour like in Fig.~\ref{fig:fig9}(a). Also for the set of nanostructures analysed here, the significant range of field-induced F interaction is present for every considered value of $J$.

\begin{figure}[t]
  \begin{center}
\includegraphics[scale=0.3]{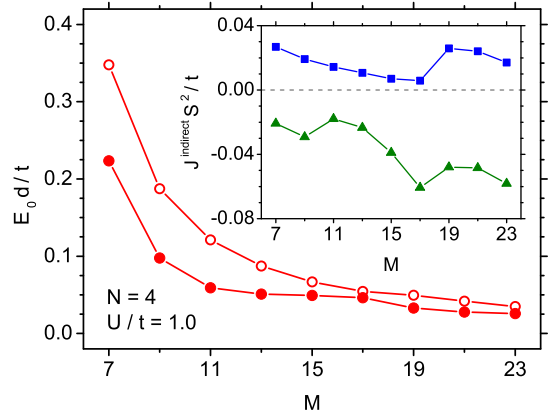}
  \end{center}
   \caption{\label{fig:fig6b}Critical normalized electric field of transition between ferro- and antiferromagnetic coupling between magnetic planes as a function of nanostructure width for fixed length $N=4$. Filled symbols denote the lower field, at which AF coupling turns into F one, while filled symbols correspond to a higher field at which opposite transition takes place. Inset presents normalized indirect coupling energy at zero field (filled rectangles) and at the first ferromagnetic maximum (filled triangles) as a function of nanostructure width for fixed length $N=4$. }
\end{figure}

The critical fields for switching between ferro- and antiferromagnetic indirect coupling can also be followed as a function of width of the mediating graphene nanostructure. The results of such calculations are presented in Fig.~\ref{fig:fig6b} for fixed length $N=4$ and for $J/t=0.1$. Like in Fig.~\ref{fig:fig9}, the filled symbols correspond to AF-F switching when electric field increases, while the empty ones denote further AF-F transition. When the width of the nanostructure increases, a monotonous decrease of the critical fields is visible, together with significant reduction of the range of fields in which F indirect coupling is present; however, such a range was always found. The inset in Fig.~\ref{fig:fig6b} presents also the indirect coupling values achieved either at zero field (strongest AF interaction) or at maximum of F coupling (at the field in between critical fields plotted in Fig.~\ref{fig:fig6b}). For these quantities, a sort of non-monotonous behaviour is found, with reduction of magnitude of F coupling when $M$ increases, while AF interaction shows rather an increase with some oscillatory tendency. Nevertheless, a considerably high sensitivity of $J^{indirect}$ to external field is present in the whole range of studied widths, and the widest structures offer rather weak critical field for switching, still with significant amplitude of both AF and F interaction.

\begin{figure}[t]
  \begin{center}
\includegraphics[scale=0.3]{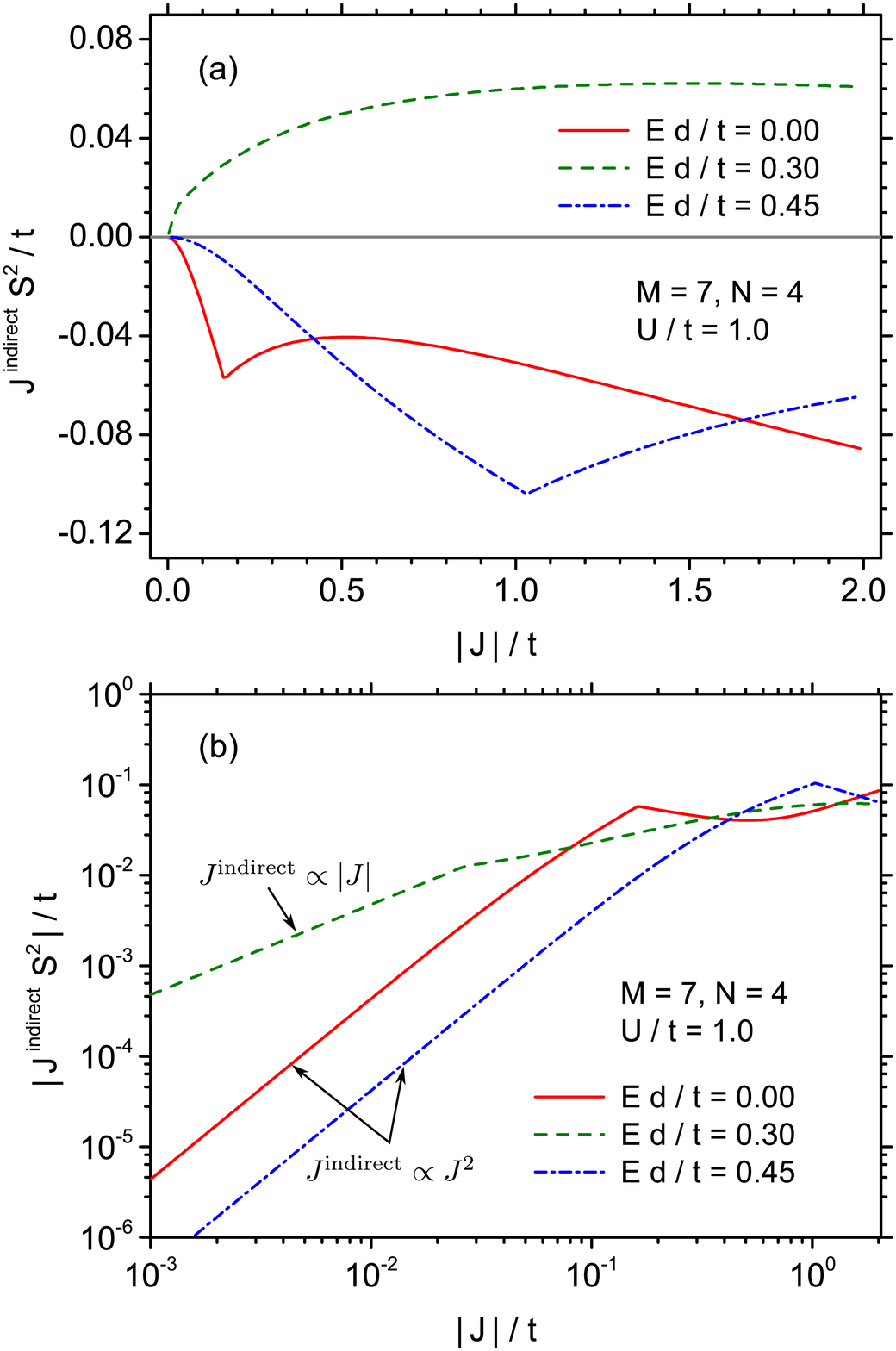}
  \end{center}
   \caption{\label{fig:fig6}Normalized indirect coupling as a function of exchange energy $J$ for three values of electric field, corresponding approximately to extremum magnitudes of indirect coupling. The results concern the nanostructure with $M=7$ and $N=4$. (a) indirect coupling in linear scale; (b) indirect coupling in double logarithmic scale.}
\end{figure}

Let us now focus the attention on the indirect interaction energy at the extremal values, i.e. for $Ed/t=0.30$ (approximately the ferromagnetic peak), for $Ed/t=0.45$ (the subsequent antiferromagnetic minimum) as well as at $Ed/t=0.0$. The dependence of $J^{indirect}$ at these field values on exchange energy $J$ is presented in Fig.~\ref{fig:fig6}(a) in linear scale and Fig.~\ref{fig:fig6}(b) in double logarithmic scale (where the absolute values are plotted). It is seen that in absence of electric field $J^{indirect}$ is proportional to $J^2$ for $J/t\lesssim 0.15$, what indicates that the origin of the indirect coupling is the second-order perturbative mechanism. The similar property is observed for coupling at $Ed/t=0.45$, i.e. at the antiferromagnetic minimum. For stronger exchange $J$, the antiferromagnetic coupling changes abruptly the slope. For $Ed/t=0$, after some range of weak dependence on $J$, it starts again to become stronger in the vicinity of $J/t\simeq 0.5$, but with reduced slope in comparison to the range of small $J$. The behaviour of $J^{indirect}$ at ferromagnetic maximum ($Ed/t=0.3$) is, however, different. For weak $J/t\lesssim 0.025$ it varies linearly with $J$, what proves that the leading contribution to the indirect coupling comes from first-order perturbative process. Such a situation is expected on the grounds of the predicted spin polarization emerging in the range of electric fields close to $Ed/t=0.3$ [see Fig.~\ref{fig:fig3}(a)]. For stronger exchange parametrized by $J$, the slope of $J^{indirect}$ is reduced and it varies with $J$ approximately proportionally to $J^{1/2}$.

The first-order perturbative contribution to indirect coupling can be described quantitatively by referring to the corrections to total energy of the system of charge carriers introduced by interaction with magnetic planes. This can be done in a very similar way as in Ref.~\onlinecite{Szalowski2013a}. Therefore, $J^{indirect}\propto \Delta E^{AF}-\Delta E^{F}$, where $\Delta E^{F,AF}$ is the first-order perturbative correction to the ground state energy of the charge carriers due to their interaction with magnetic planes (with either F or AF orientation of magnetic moments). This correction is directly proportional to the total spin density at the appropriate sites where the contact interaction occurs. 

\begin{figure}[t]
  \begin{center}
\includegraphics[scale=0.3]{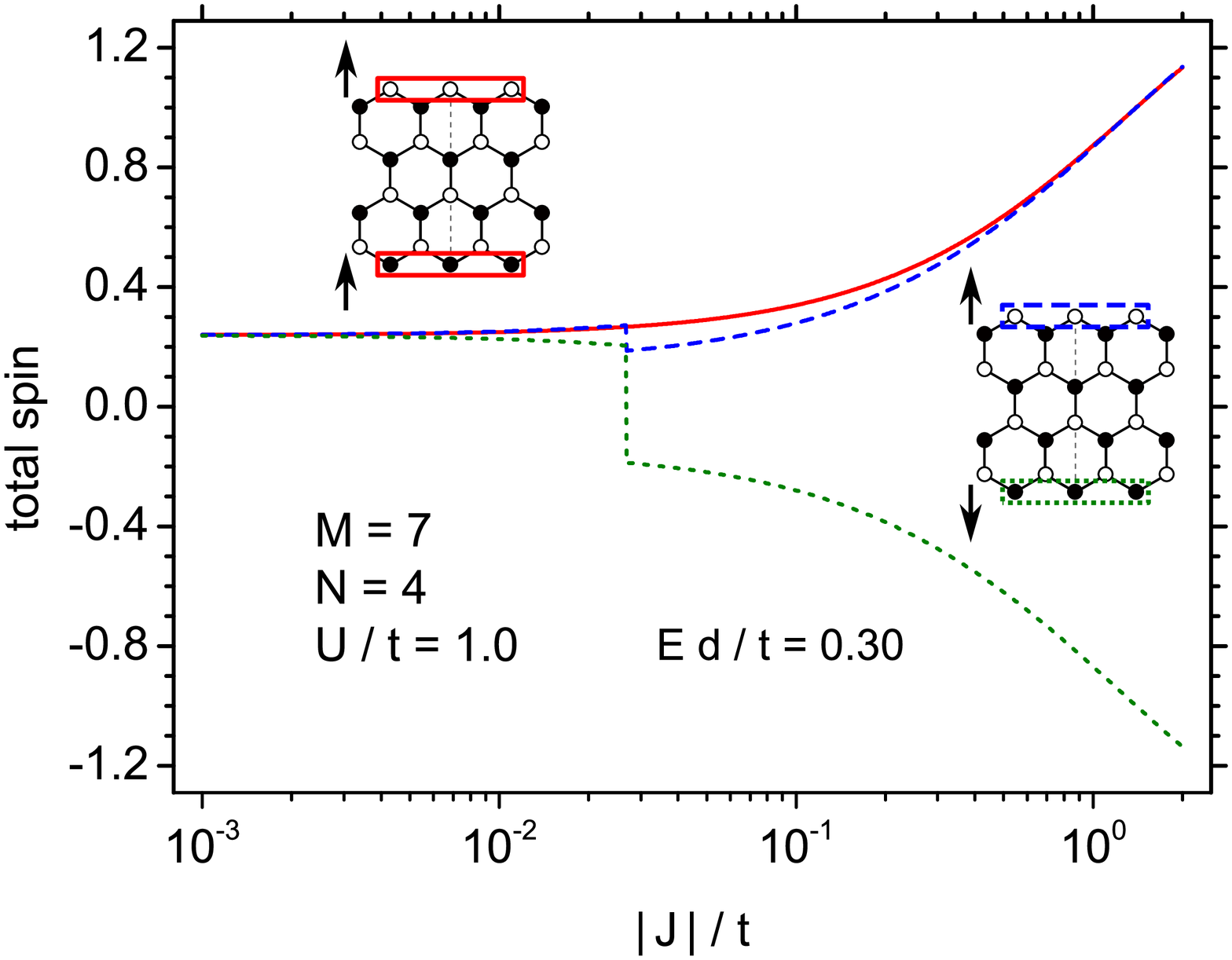}
  \end{center}
   \caption{\label{fig:fig7} Total spin density at outermost sites of zigzag terminations of the nanostructure with $M=7$ and $N=4$ as a function of exchange energy $J$, for ferromagnetic (solid line) and antiferromagnetic (dashed and dotted lines) orientation of magnetic moments in both planes attached to the nanostructure. Electric field of $Ed/t=0.30$ is present.}
\end{figure}

In Fig.~\ref{fig:fig7} we present the dependence of the total spin density at sites interacting with magnetic planes (separately for both nanostructure terminations) on exchange energy $J$ in the presence of electric field $Ed/t=0.30$. The outermost sites are schematically marked with frames in the insets. The cases of plane magnetizations oriented ferromagnetically and antiferromagnetically are considered separately, as indicated in the insets. If magnetizations of both planes are parallel, the total spin density at both nanostructure terminations takes identical values. It is visible that this value increases monotonously with $J$; the rise is very slow for low $J/t$ and for $J/t\gtrsim 0.2$ the slope increases and $s\propto J^{1/2}$. The situation is somehow different when magnetizations of both planes are antiparallel. In such circumstances the total spins at both terminations are different. For $J/t\lesssim 0.025$, at one termination the spin rises with $J$, while at the other one it decreases, while both spins keep parallel orientation. However, at $J/t\simeq 0.025$ the spins become antiparallel and their absolute values become identical (a kind of metamagnetic transition takes place). Therefore, the total spin density at outermost atoms of both zigzag terminations is equal to 0. It can be concluded that for $J/t\lesssim 0.025$ the spin density is only weakly influenced by exchange interaction with magnetic planes. On the other hand, for $J/t\gtrsim 0.025$, the leading contribution to indirect coupling originates only from the ferromagnetic state of magnetization of both planes. Therefore, since spin density approximately increases according to $s\propto J^{1/2}$ , also $J^{indirect}$ follows this proportionality. Let us state that this is no longer a first-order contribution to indirect coupling, since the spin density is modified by exchange interaction between planar magnetic moments and spins of charge carriers and such a modified value enters the formula for indirect exchange. The second one of the mentioned ranges seems to contain the most important values of $J$ for practical uses.

\begin{figure}[t]
  \begin{center}
   \includegraphics[scale=0.3]{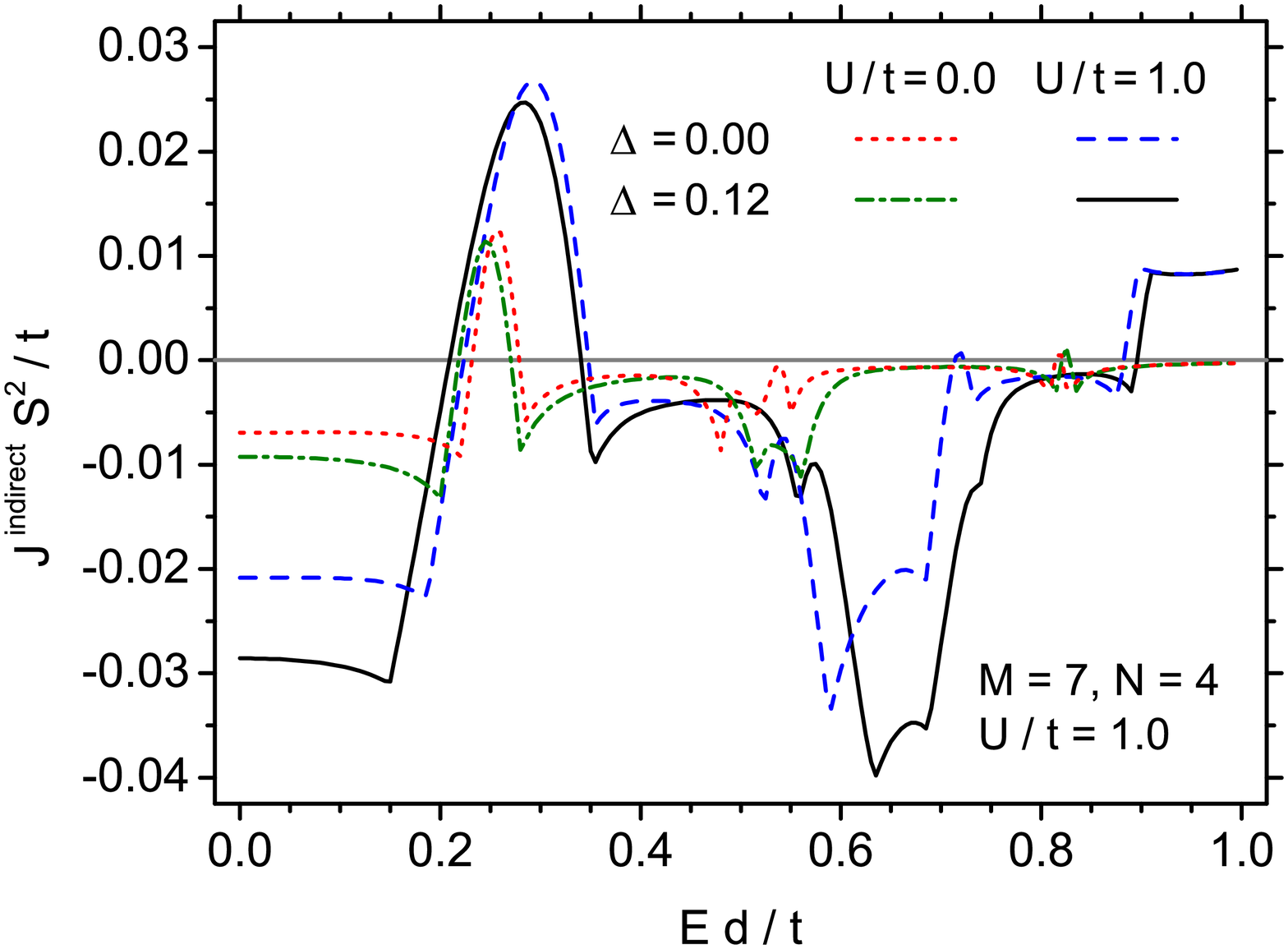}
  \end{center}
   \caption{\label{fig:fig8}Indirect coupling between the planes as a function of normalized external electric field for a nanostructure with $M=7$ and $N=4$, in presence and in absence of coulombic interactions (parametrized by Hubbard parameter $U$) and armchair edge deformation of the nanostructure.}
\end{figure}

In order to identify the crucial factors shaping the response of the indirect coupling to the electric field, let us supplement the previous results with those shown in Fig.~\ref{fig:fig8}. In that plot, $J^{indirect}$ is shown as a function of normalized electric field for the representative and most interesting nanostructure with $M=7$ and $N=4$. Different lines correspond to the results in which armchair edge deformation (parametrized by $\Delta$) and coulombic interactions (parametrized by Hubbard $U$ parameter) are separately included or excluded. First it is noticeable that the weakest antiferromagnetic indirect exchange at zero electric field occurs for the absence of coulombic interactions and edge deformation. Emergence of the edge bonds shortening for $U/t=0$ is only weakly reflected in $J^{indirect}$ , while switching on $U$ results in considerable strengthening of indirect coupling. For the range of electric fields for which switching to ferromagnetic interaction happens, the similar situation can be seen. Namely, the most crucial factor influencing the coupling is coulombic interaction (positive $U$ value), while $\Delta$ has only slight effect on $J^{indirect}$. If the electric field becomes stronger, the important influence of edge deformation can be noticed mainly in presence of $U/t=1.0$ for the deep antiferromagnetic minimum. It can be generally concluded that the presence of coulombic interaction tends to boost the magnitude of indirect coupling, conserving its sign. This factor also extends significantly the range of ferromagnetic interaction peak present at $Ed/t\gtrsim 0.2$, shifting the field for which sign changes back to antiferromagnetic one towards higher values. Therefore, coulombic interactions promote the features which are desirable from the point of view of obtaining a clearly field-controlled indirect coupling. Let us also observe that the electric field switching of $J^{indirect}$ between antiferro- and ferromagnetic values is observable for every choice of parameters in Fig.~\ref{fig:fig8}, what allows to state that this feature is considerably robust.

\section{Final remarks}

In our study we have demonstrated the prediction of indirect coupling between magnetic planes mediated by charge carriers in a graphene nanostructure of particular size and geometry. The crucial characteristic of this interaction is its sensitivity to external electric field in plane of the nanostructure. Namely, the coupling which is antiferromagnetic below certain critical electric field can be continuously switched to ferromagnetic sign for field strong enough. Further increase of the field brings the coupling back to antiferromagnetic regime. The effect is present for a wide range of exchange energies between attached magnetic planes and spins of the charge carriers. Moreover, it is also robust against deformation of the armchair edge of the mediating nanostructure and boosted by coulombic interactions. The origin of the effect is connected with field-induced appearance of the nonzero spin density distributed along the zigzag termination, not necessarily connected with net magnetic polarization of the structure as a whole. The extreme values of ferro- and antiferromagnetic indirect interaction are found to be of comparable magnitude. 

Let us emphasize that the nanostructures mediating the interaction (in the form considered in the present study) appear experimentally accessible and their magnetic properties can be examined \cite{Konishi2013}. Moreover, there is also a constant progress in achieving well-controllable graphene nanoribbons, including the development of bottom-up methods utilizing molecular precursors (see e.g. \cite{Cai2010,Yazyev2013}). This encourages studies of geometrically well-defined graphene structures, focusing on their usefulness in spintronics. Moreover, in presence of magnetic planes attached to the zigzag edges of the nanostructure, the conditions for emergence of zigzag edge spin polarization should be more favourable than those in free-standing graphene, since the magnetic planes serve as a source of exchange field and magnetic anisotropy stabilizing the ordering against thermal excitations. This factor, together with recent results concerning zigzag edge magnetism \cite{Joly2010,Konishi2013b}, supports interest in such systems.

The described effect can find application in construction of electric-field controlled spintronic devices based on graphene. The further developments may include search for analogous field-tunable effects among structures of various geometry as well as, for example, study of simultaneous effect of electric and magnetic field on the described phenomenon. Moreover, the effect of charge doping might be worthy of investigations. However, let us mention that the properties of the smallest mediating nanostructures appear the most promising, since the increase in size causes inevitably the screening effects to reduce the influence of external electric field on the desirable properties \cite{Yamanaka2012,Yamanaka2014}. Let us also comment that for the purpose of increasing the energy of indirect coupling between the magnetic planes, the  device would be imagined with numerous identical mediating nanostructures attached to the planes. These structures would be placed at equal distance one from another, possibly each structure accompanied with its own pair of electric gates. In such a way the coupling would be enhanced.

\section{Acknowledgments}

\noindent The computational support on Hugo cluster at Laboratory of Theoretical Aspects of Quantum Magnetism and Statistical Physics, P. J. \v{S}af\'{a}rik University in Ko\v{s}ice is gratefully acknowledged.

\noindent This work has been supported by Polish Ministry of Science and Higher Education on a special purpose grant to fund the research and development activities and tasks associated with them, serving the development of young
scientists and doctoral students.

\end{document}